\documentclass[letterpaper]{article}
\usepackage{times}
\usepackage{latexsym}
\usepackage{graphicx}
\usepackage{mathtools}
\usepackage{amsmath}
\usepackage{enumitem}
\usepackage{wrapfig}
\usepackage{caption}
\usepackage{authblk}
\usepackage{subcaption}
\usepackage[utf8]{inputenc}
\usepackage[colorlinks]{hyperref}       
\usepackage[margin=1.25in]{geometry}

\title{Learning Noise-Invariant Representations \\
for Robust Speech Recognition}
\author{Davis Liang$^1$, 
Zhiheng Huang$^1$,
Zachary Lipton$^{1,2}$ \\
$^1$Amazon AI\\
$^2$Carnegie Mellon University\\
\href{mailto:liadavis@amazon.com}{\nolinkurl{liadavis@amazon.com}},
\href{mailto:zhiheng@amazon.com}{\nolinkurl{zhiheng@amazon.com}},
\href{mailto:zlipton@cmu.edu}{\nolinkurl{zlipton@cmu.edu}}}

\usepackage{booktabs}
\usepackage{array}
\newcolumntype{L}[1]{>{\raggedright\arraybackslash}p{#1}}
\usepackage{xcolor}
\date{}

\begin{document}
\maketitle

\begin{abstract}
Despite rapid advances in speech recognition, 
current models remain brittle 
to superficial perturbations to their inputs. 
Small amounts of noise can destroy the performance 
of an otherwise state-of-the-art model. 
To harden models against background noise, 
practitioners often perform data augmentation, 
adding artificially-noised examples 
to the training set, carrying over the original label. 
In this paper, we hypothesize 
that a clean example and its superficially perturbed counterparts shouldn't merely map to the same \emph{class}
--- they should map to the same \emph{representation}. 
We propose invariant-representation-learning (IRL): 
At each training iteration, for each training example,
we sample a noisy counterpart. 
We then apply a penalty term 
to coerce matched representations at each layer 
(above some chosen layer). 
Our key results, demonstrated on 
the Librispeech dataset are the following: 
(i) IRL significantly reduces character error rates (CER)
on both `clean' ($3.3\%$ vs $6.5\%$) and `other' ($11.0\%$ vs $18.1\%$) test sets; 
(ii) on several out-of-domain noise settings 
(different from those seen during training), 
IRL's benefits are even more pronounced. 
Careful ablations confirm that 
our results are not simply due to 
shrinking activations at the chosen layers. 
\end{abstract}

\section{Introduction}
\label{sec:intro}
Over the past several years, 
a series of papers have developed end-to-end deep learning systems 
for automatic speech recognition (ASR), 
advancing the state of the art on a variety of benchmarks 
\cite{hannun2014deep, dario2014deep, miao2015eesen, 
bahdanau2016end, zeyer2018, zhou2018}.
Typically, these models consist of either 
Recurrent Neural Networks (RNNs)
with Sequence-to-Sequence (Seq2Seq) 
architectures \cite{sutskever2014sequence} 
and attention mechanisms \cite{luong2015effective, bahdanau2014}, 
RNN transducers \cite{graves2012sequence}, 
transformer networks \cite{vaswani2017attention, zhou2018syllable}, 
convolutional neural networks paired with transformer networks 
\cite {zhou2018comparison, collobert2016wav2letter}, 
or RNNs trained with CTC loss \cite{graves2013speech}. 
Often, these models act on spectral features, 
e.g., Mel-Frequency Cepstral Coefficients (MFCC) \cite{davis1980}.

While these systems achieve impressive accuracy 
when trained and evaluated on clean data, 
they suffer a well-documented sensitivity to changing noise levels 
and various noise types \cite{bhattacharjee2016}.
Perhaps this vulnerability should not be surprising,
given the significant impact that background noise 
can have on MFCC features \cite{bhattacharjee2016}. 

One simple strategy to combat the vulnerability 
of deep nets to background noise 
is a technique known generally as \emph{data augmentation}, 
and as \emph{multi-condition training} 
in the speech recognition community.
Here, we augment the original data by applying transformations 
to which we want our models to be invariant 
and assigning these perturbed data points 
the same label as the unperturbed originals.
While the computer vision literature 
has long focused on perturbations like random crops, 
rotations, translations, and Gaussian noise 
\cite{krizhevsky2012imagenet, an1996effects, 
grandvalet1995comments, bishop1995training, grandvalet1997noise},
data augmentation papers in the ASR literature commonly sample snippets of additive background noise 
from datasets such as MUSAN \cite{snyder2015},
which contains environmental noise 
(dial tones, thunder, footsteps, animal noises, etc), 
music (baroque, classical, romantic, jazz, bluegrass, hip-hop, etc.), 
and speech. 
ASR models trained with such augmented data 
have demonstrated lower grapheme error rates 
on noisy data \cite{yin2015, yu2016effect, rajnoha2009multi}.

In this paper, we draw inspiration from the human ability
to recognize not only that a clean clip 
and its noisy counterpart 
belong to the same category
but that they are produced from the same exact recording. 
Thus, we propose models 
that map both clean inputs and their noisy counterparts 
onto the same point in representation space,
introducing this inductive bias 
via regularization terms, 
penalizing differences between the hidden representations 
produced from real and noisy data.
Throughout training, for each clean example, 
we synthesize one noisy counterpart,
using a custom data augmentation pipeline
that first selects a random noise snippet and volume level, 
adding the two raw waveforms
and then generating the corresponding MFCC features on the fly.
At each iteration, we apply the original cross-entropy loss on the predictions for both clean and perturbed inputs and also penalize the difference in hidden activations encouraging corresponding activations 
as quantified by both cosine distance and $L_2$ distance.

Our experiments address the Librispeech dataset \cite{panayotov2015},
building on a Seq2Seq baseline with cross-entropy loss.
To keep the empirical study clean, 
we do not use a language model.
We run all experiments both on the standard dev and test sets 
and also under a variety of out-of-domain noise conditions.
First, we show that 
while data augmentation improves generalization error 
on both the original task and under out-of-domain noise, 
the models still suffer significant degradation 
in performance. 
Next, we show that Invariant-Representation Learners (IRLs) improve significantly over generic data augmentation models, 
both on the \emph{clean} and \emph{other} 
(the more challenging dataset with higher word error rate) subsets of the LibriSpeech test set.
Comparisons against an adversarial approach proposed by \cite{serdyuk2016} and the logit pairing approach due to \cite{kannan2018} demonstrate the significant advantage of IRL.
We then demonstrate that on a variety of simulated out-of-domain noise conditions, the IRL models are 
considerably more robust than all baselines.
Finally, we perform ablation experiments, 
showing that our models trained with the IRL algorithm outperform 
well-known regularization tactics 
like weight decay applied on the same representations.

\subsection{Related Work}
\label{sec:relatedwork}
A number of proposed models address the goal of 
noise-robust speech recognition: 
\cite{seltzer2013investigation} 
proposes a method called \emph{noise-aware training}
that introduces information about the environment 
as additional inputs to DNN-based acoustic models. 
\cite{yin2015} proposes augmenting training examples 
with additive noise sampled 
from the DEMAND noise database  training examples. 
\cite{serdyuk2016} seeks noise-invariant representations in  DNN-HMM architectures through an adversarial learning setup.
\cite{huang2013audio} shows the training on multi-modal data leads to noise robust models.
\cite{gemmeke2011exemplar} demonstrates 
that modeling speech as a linear combination of exemplars results in noise-robust ASR models.
\cite{maas2012recurrent} proposes deep recurrent autoencoders 
to denoise input features. 
\cite{li2014overview} presents an overview of methods 
for noise-robust ASR, including recursive cepstral mean and variance normalization \cite{viikki1998recursive}, 
joint adaptive training \cite{liao2007adaptive}, 
and speaker adaptive training \cite{anastasakos1996compact}. 
To our knowledge, no prior work in speech recognition
employs our simple approach of penalizing 
distance between the hidden representations corresponding 
to clean and noisy signals.

In the most similar paper, \cite{serdyuk2016} claimed 
that with adversarially trained DNN-HMM systems, 
the best performance gain is achieved 
when a small number of noise types are available for training. When using $6$ noise classes (airport, babble, car, restaurant, street, and train), 
\cite{serdyuk2016} found that there was 
no significant difference between 
the adversarial and baseline models. 
In contrast, our models show a CER improvement over baseline of 3.1\% absolute on test-clean and 6.5\% absolute on test-other using hundreds of noise classes.

\section{Noise-Invariant Representations}
\label{sec:models}
To begin, we formally describe our loss function
for enforcing noise-invariant representations 
on the outputs of a given layer.
Because our first proposed model focuses noise-invariance 
in the encoding layer,
we dub models using such loss functions \emph{IRL-E}. 
In other experiments, we apply a cumulative penalty, 
additionally requiring noise-invariant representations at all subsequent layers, naming this model \emph{IRL-C}.
We begin by describing IRL-E. 
Subsequently, extension to IRL-C will be straightforward.

\subsection{IRL-E}

\begin{figure}[t]
  \centering
  \includegraphics[keepaspectratio, width=0.5\textwidth]{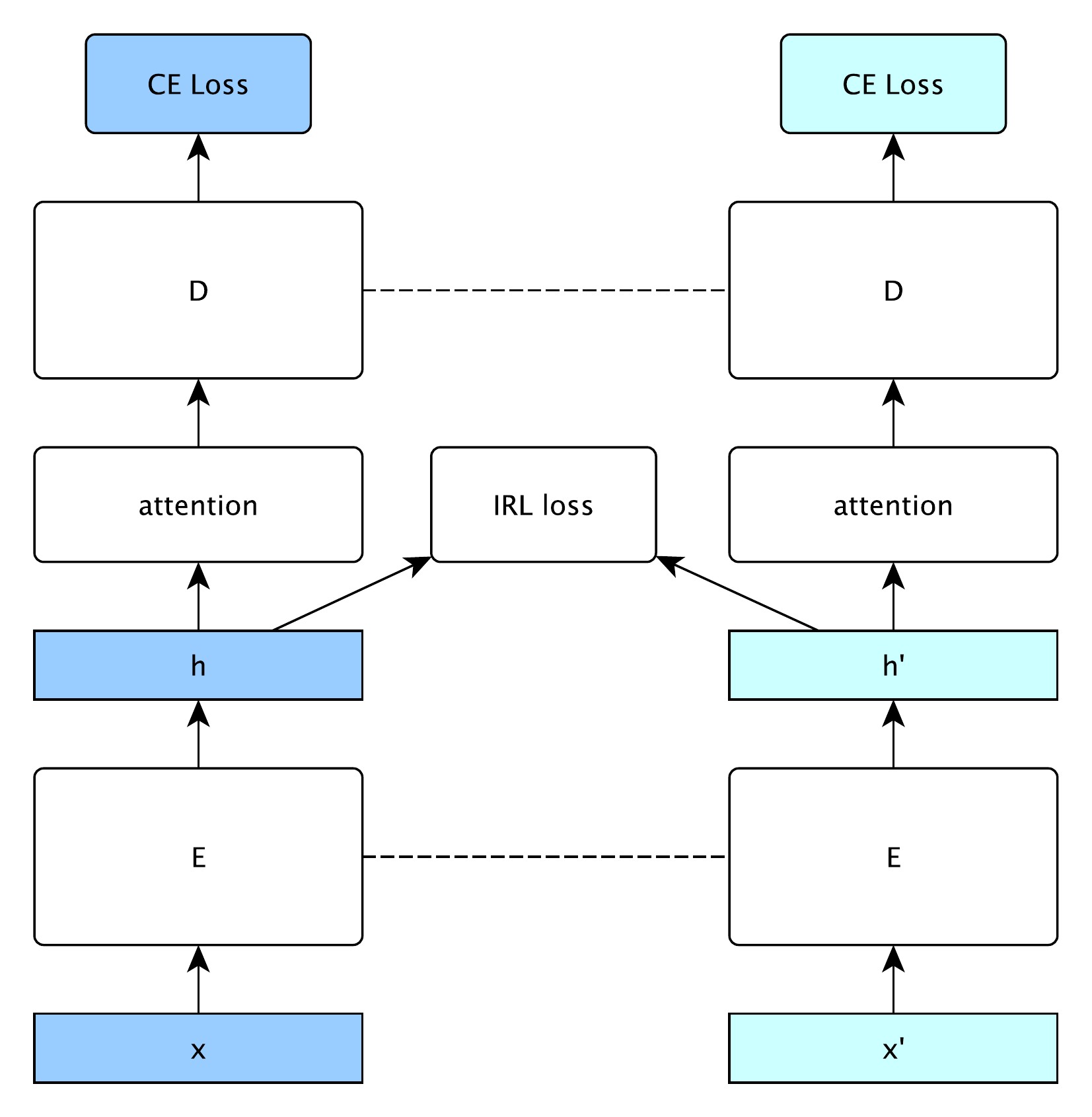}
  \caption{Diagram demonstrating the various terms in the IRL loss function as applied to a Seq2Seq attention model. Dotted lines represent shared weights.}
  \label{fig:domaininvariantlossmodel}
\end{figure}

The IRL algorithm is simple: First, during training, 
for each example $\mathbf{x}$, 
we produce a noisy version by sampling from $\mathbf{x}' 
\sim \nu(\mathbf x)$,
where $\nu$ is a stochastic function. 
In our experiments, this function 
takes a random snippet from a noise database, 
sets its amplitude by drawing from a normal distribution,
and adds it to the original (in sample space), 
before converting to spectral features.
We then incorporate a penalty term in our loss function
to penalize the distance between the encodings of the original data point $\phi_e(\mathbf x)$ and the noisy data point $\phi_e(\mathbf x')$,
where $\phi_{l}$ is representation at layer $l$.
In our experiments, we choose $\phi_e$ to be the output 
of the encoder in our Seq2Seq model.
We illustrate the learning setup graphically in Figure \ref{fig:domaininvariantlossmodel}.
In short, our loss function consists of three terms,
one to maximize the probability assigned to the the clean example's label,
another to maximize the probability our model assigned to the noisy example's (identical) label scaled by hyper-parameter $\alpha$, 
and a penalty term to induce noise-invariant representations $L_d$. In the following equations, we express the loss calculated on a single example $\mathbf{x}$ and its noisy counterpart $\mathbf x'$,
omitting sums over the dataset for brevity.
\begin{equation*}
L(\theta) = L_c(\mathbf{x}; \theta) + \alpha L_c(\mathbf{x}'; \theta) + L_d (\mathbf{x}, \mathbf x'; \theta ),
\end{equation*}
where $\theta$ denotes our model parameters.
Because our experiments address multiclass classification, 
our primary loss $L_c$ is cross-entropy:
\begin{equation*}
L_c(\mathbf x; \theta) = - \sum_{k=1}^{C} y_{k} \log \hat{y}_{k}(\mathbf{x}; \theta),
\end{equation*}
where C denotes the vocabulary size and $\hat{y}$ is our model's softmax output. 
To induce similar representations for clean and noised data, 
we apply a penalty consisting of two terms, 
the first penalizes the $L_2$ distance 
between $\phi_e(\mathbf x)$ and $\phi_e(\mathbf{x}')$,
the second penalizes their negative cosine distance.
\begin{align*}
L_d( \mathbf{x}, \mathbf x'; \theta) =& \gamma  \left( \phi_e(\mathbf{x})- \phi_e(\mathbf x') \right)^{2} - \lambda \frac{\phi_e(\mathbf{x}) \cdot \phi_e(\mathbf x')}{||\phi_e(\mathbf{x})|| \cdot ||\phi_e(\mathbf x')||}
\end{align*}

We jointly penalize the $L_2$ and cosine distance 
for the following reason. 
It is possible to lower the $L_2$ distance 
between the two (clean and noisy) hidden representations 
simply by shrinking the scale of all encoded representations. 
Trivially, these could then be dilated again 
simply by setting large weights in the following layer.  
On the other hand, it is possible to assign high cosine similarity to the two vectors but for their magnitudes to vary significantly. 
By jointly penalizing $L_2$ and cosine distance, 
we require that both the clean and noisy representations point in the same direction and are close to each other in magnitude.

\subsection{Applying IRL Cumulatively Across Layers (IRL-C)}
It is possible for representations 
to be close (but not identical) in the encoder layer,
but to subsequently be pushed apart 
in subsequent decoder layers. 
Thus, we introduce another model, 
\emph{IRL-C} (C for \emph{cumulative}), 
that additionally applies the IRL penalty on all 
subsequent decoder layers. 
By requiring noise-invariant representations in multiple layers, 
we ensure that each training example 
and its randomly-sampled noisy counterpart 
have similar representations throughout the network.
Note that if the encodings of the clean and noisy examples 
are identical at the encoder layer, 
then all subsequent layers will also be identical 
and thus those penalties will go to $0$.
We can express this loss as a sum 
over successive representations $\phi_l$ 
of the clean $\phi_{l}(\mathbf x)$ 
and noisy $\phi_{l}(\mathbf x')$ data:
\begin{align*}
L_d( \mathbf{x},  \mathbf x'; \theta ) =& \sum_{l=e}^{L} \Bigg[ \gamma ( \phi_l(\mathbf{x})- \phi_l(\mathbf x') )^{2} - \lambda \frac{\phi_l(\mathbf{x}) \cdot \phi_l(\mathbf x')}{||\phi_l(\mathbf{x})|| \cdot ||\phi_l(\mathbf x')||} \Bigg]
\end{align*}
In our experiments, we find that IRL-C consistently gives a small improvement over results achieved with IRL-E.

\subsection{Application to Recurrent Speech Models}

As described to this point, our loss can be applied on any feedforward neural network with any noise process $\nu$.
Applying our technique to recurrent neural networks requires just a few additional considerations.
Primarily, we must decide how to deal with the 
sequence structure. 
Two natural choices are (i) to concatenate the representations for a given layer across time steps, and then to apply our penalty on the concatenated representations and 
(ii) to apply the penalty separately at each time step and then to sum (or equivalently, up to a scaling factor to average) 
over the time steps.
These approaches are identical for the $L_2$ penalty 
but not for the cosine distance penalty, 
owing to the normalizing factor 
which may be different at each time step.
In this work we take approach 
(i) concatenating the representations across time steps 
and then calculating the penalty.

All of our models are based off of the sequence-to-sequence 
due to \cite{bahdanau2014}. 
The input to the encoder is a 
sequence of spectral features, here MFCC, 
which are encoded by several consecutive layers of LSTM units. The encoder output states are then passed through an attention mechanism which computes the similarity between the decoder hidden states and the encoder output states. 
The output is a softmax over the vocabulary (here, characters) at each decoder time step. 

In our experiments with IRL-E 
(penalty applied on a single layers),
we use the output of the encoder to calculate the penalty. 
Note that there is one output per step in the input sequence and thus we are concatenating across the $T_1$ steps. 

To calculate IRL-C, we also start with the encoder output 
concatenating across all $T_1$ sequence steps 
to calculate the IRL penalty.
However, for all subsequent layers, 
we are acting upon layers in the decoder,
and thus concatenating across 
the number of decoding sequence steps $T_2$ 
for calculating these terms in the IRL-C penalty.


\section{Datasets}
\label{sec:datasets}
\paragraph{Librispeech}
We evaluate all models on 
the LibriSpeech \cite{panayotov2015} dataset. 
This dataset consists of roughly $1000$ hours of audio split into training, dev and test partitions.
The dataset was carefully designed 
to ensure that no speaker (person) appears 
in multiple partitions. 
Within both the dev and the test partitions, 
the data is further subdivided 
into ``clean'' and ``other'' subsets based on the speakers. 
The ``clean'' portion contains those speakers for which a baseline model had the lowest CER,
and the ``other'' portion contains those speakers for whom the error rate was high.
Following common practice in the literature on these datasets,
we evaluate all models on the dev-clean, dev-other, test-clean, and test-other splits separately. 

\paragraph{The MUSAN Noise Dataset}
For our additive noise, we draw upon samples 
from the MUSAN noise dataset \cite{snyder2015}. 
MUSAN was released under a flexible Creative Commons license and consists of approximately $109$ hours 
of noise sampled at $16$kHz. 
The dataset contains music from several genres, namely baroque, classical, romantic jazz, bluegrass, and hip-hop, among others, 
speech from twelve languages, 
and a wide assortment of technical and non-technical noises.
To generate noisy audio, 
we first add MUSAN noise to the training data point at a signal-to-noise ratio drawn from a Gaussian with a mean of $12$dB and a variance of $8$dB.
This aligns roughly with the scale of noise 
employed in other papers using multi-condition training
\cite{dario2014deep}.

\section{Experiments}
\label{sec:experiments}
Before presenting our main results, 
we briefly describe the model architectures, training details, and the various baselines 
that we compare against. 
We also present details on our pipeline  
for synthesizing noisy speech 
and explain the experimental setup 
for evaluating on out-of-domain noise.

\subsection{Model Architecture}
To facilitate reliable comparisons between 
our methods and various baseline training schemes,
we conduct all experiments using 
identical architectures and tuning schemes.
Because we conduct a large number of experiments 
and because of the computational expense 
of unrolling of long speech sequences, 
we struck a balance between performance and speed
when choosing the basic architecture.
The encoder for our base model consists of $4$ layers: 
$2$ encoder BLSTM layers with $320$ hidden units each,
followed by $2$ encoder LSTM layers with $320$ hidden units each.
Our decoder accesses the encoded representations 
using dot product attention, 
and contains $4$ decoder LSTM layers, 
with $320$ hidden units each. 
Notably, our first encoder layer 
halves the sequence length 
by concatenating adjacent inputs along the temporal axis. 
Each model across all of our comparisons 
has the exact same number of trainable parameters. 
To keep things simple, we do not use an external language model.
Instead we decode predictions from all models
via beam search with width $10$. 

To ensure fair comparisons, 
we perform hyper-parameter searches separately for each model 
and account for variability due to initialization 
by training each model 5 times and keeping the best run 
as determined on the dev-other partition. 
Specifically, we tune the weights on our losses 
by trying each of the scale values 
($0.001$, $0.01$, $0.1$, $1$, $10$, and $100$). 
We found that an $\alpha$ of $1$ 
(the weight on the cross-entropy loss of the noised data), 
a $\gamma$ of $0.01$ (the weight on the L2 distance loss), 
and a $\lambda$ of $0.01$ (the weight on the cosine distance loss) 
worked well.


\subsection{Training Details}
We train all models with the Adam optimizer 
with an initial learning rate of $0.001$. 
We employ a learning rate schedule similar to NewBob \cite{quicknet} 
that decreases by a factor of $2$ 
if there is an increase in validation perplexity epoch-over-epoch. 
We employ a stopping criterion 
that ends training if validation perplexity 
does not decrease for three epochs in a row. 
We limit each models to a maximum of $40$ epochs, 
although our networks generally converge within $20$ epochs.

The primary loss function for each model
is cross-entropy loss and our primary evaluation metric
to evaluate all models
is the character error rate. 
As described above, the additional loss terms 
for our IRL models are L2 loss and cosine distance
between representations of clean and noisy audio.

\subsection{Baselines}
Our baseline models include a model trained 
on the standard training data, 
a model trained with noise-augmented data, 
a model trained with noise augmented data and weight decay, 
and a data augmented model supervised with L2 loss 
to push activations of the encodings to $0$.  
These ablation tests provide evidence 
that our IRL algorithm isn't 
simply penalizing the norm of the encodings.

\begin{itemize}
\item \textbf{Baseline: } 
Our base model trains the baseline sequence-to-sequence model on the original $960$ hours of Librispeech training data.
\item \textbf{Data Augmentation: } Our data augmentation model trains the sequence-to-sequence model on both the examples from the 960 hour Librispeech training corpus and the randomly generated noisy counterparts. 
\item \textbf{Adversarial: } 
The adversarial model consists of an adversarial noise discriminator 
trained on top of the encoder outputs. 
The discriminator consists of $2$ layers of $256$ ReLu units 
and a single unit sigmoid output.
We train the discriminator to classify 
whether the representation originates from clean or noised inputs. 
The encoder meanwhile is trained both to minimize the classification loss
and to fool the discriminator, in a scheme similar to the reverse gradient technique in the domain-adversarial approach due to \cite{ganin2016domain}
and applied to speech by \cite{serdyuk2016}.
\item \textbf{Logit Pairing: }
Our final baseline consists of the logit pairing model due to \cite{kannan2018} which applies L2 loss and cosine distance loss 
on the final decoder layer logits, 
enforcing noise-invariant representations but only on the output layer. 

\end{itemize}

\subsection{Synthesizing Noise}
We train all models on the LibriSpeech corpus,
generating noisy data by adding randomly selected noise tracks 
from the MUSAN dataset 
with a signal to noise ratio 
drawn from a Gaussian distribution 
($12$dB mean, $8$dB standard deviation) 
and temporal shift drawn from a uniform distribution 
(with a range of $0$ to $1000$ms). 
For the data augmentation model, 
this result resembles the typical 
data augmentation (multi-condition training) procedure. 

\subsection{Out-of-Domain Noise}
Next, we evaluate each of our models 
on a variety of noise conditions 
that were not seen at training time. 
In particular, we consider the following out-of-domain noise conditions:
(i) augmenting the test-clean split 
with overlapping out-of-domain speech from the WSJ-0 dataset \cite{wsj} to simulate multi-speaker environments,
(ii) applying additive noise 
with various SNRdb to simulate varying noise levels, 
(iii) modulating the volume of the clean signal 
to simulate different levels of speaker loudness, 
(iv) convolving the original wave file 
with room impulse responses to simulate the effect 
of room reverberation on speech, 
and (v) re-sampling to $8$kHz to simulate telephoney data. 
For each setting, we measure CER on the out-of-domain
noise-augmented test-clean data.

\section{Results}
\label{sec:results}
\begin{table*}[!htbp]
\centering
\begin{tabular}
{@{}p{0.15\textwidth}*{4}{L{\dimexpr0.2\textwidth-2\tabcolsep\relax}}@{}}
\toprule 
& \multicolumn{2}{c}{\bf{Evaluation Set}} &
\multicolumn{2}{c}{\bf{Test Set}} \\
\cmidrule(r{4pt}){2-3} \cmidrule(l){4-5}
& dev-clean & dev-other & test-clean & test-other\\
\midrule
Baseline & $6.7$\% & $17.8$\% & $6.5$\% & $18.1$\% \\
Data Aug. & $6.4$\% & $16.8$\% & $6.4$\% & $17.5$\% \\
Adversarial & $6.7$\% & $16.7$\% & $6.5$\% & $17.6$\% \\
Logit Pairing & $5.1$\% & $14.5$\% & $5.1$\% & $14.8$\% \\
IRL-E & $3.6$\% & $11.0$\% & $3.5$\% & $11.2$\% \\
IRL-C & $\bf{3.4}$\% & $\bf{10.7}$\% & $\bf{3.3}$\% & $\bf{11.0}$\% \\
\bottomrule
\end{tabular}
\caption{Evaluation and Test Set Character Error Rate on the Librispeech Corpus.}
\label{tab:main_accuracy}
\end{table*}

\begin{table*}[t]
\centering
\begin{tabular}{@{}p{0.2\textwidth}*{6}{L{\dimexpr0.12\textwidth-1\tabcolsep\relax}}@{}}
\toprule
& \multicolumn{6}{c}{\bf{CER on Noisy Data}}\\
\cmidrule(r{4pt}){2-7}
& Base & Data Aug. & Adv. & Logit & IRL-E & IRL-C\\
\midrule
Error on test-clean & $6.5$\%  & $6.4$\%  & $6.5$\% & $5.1$\% & $3.5$\% & $\mathbf{3.3}$\% \\
In-domain ($6$SNRdB)  & $27.8$\% & $10.8$\% & $16.5$\% & $8.7$\% & $6.0$\% & $\mathbf{5.7}$\% \\
In-domain ($12$SNRdB) & $13.5$\% & $7.8$\%  & $12.1$\% & $6.2$\% & $4.2$\% & $\mathbf{4.1}$\% \\
\midrule
Impulse Convolve    & $24.1$\% & $21.0$\% & $28.3$\% & $47.6$\% & $18.0$\% & $\mathbf{13.8}$\% \\
Speech ($6$SNRdB)     & $91.5$\% & $32.0$\% & $67.7$\% & $33.0$\% & $16.4$\% & $\mathbf{14.1}$\% \\
Speech ($12$SNRdB)    & $77.8$\% & $15.2$\% & $34.7$\% & $11.1$\% & $7.6$\% & $\mathbf{6.8}$\% \\
Volume ($+6$ dB)       & $6.5$\%  & $6.4$\%  & $9.8$\% & $5.1$\% & $3.6$\% & $\mathbf{3.5}$\% \\
Volume ($-6$ dB)       & $6.5$\%  & $6.3$\%  & $9.6$\% & $5.0$\% & $3.6$\% & $\mathbf{3.5}$\% \\
Telephoney          & $14.2$\% & $12.2$\% & $21.3$\% & $10.3$\% & $7.1$\% & $\mathbf{6.4}\%$ \\
\bottomrule
\end{tabular}
\caption{Character error rate for test-clean augmented with noise}
\label{tab:outofdomain}
\end{table*}

Our IRL-C model achieves the best CER 
on both test-clean and test-other 3.3\% and  11\%, respectively (Table \ref{tab:main_accuracy}). 
This compares baseline scores of $6.5\%$ and $18.1\%$, respectively. 
We note that by comparison, 
conventional data augmentation is only marginally effective.
Among the baselines that we consider, logit pairing performs best (5.1\% and 14.8\%) although the improvements are not comparable to either IRL model.

We found that weight decay slowed down network convergence and did not outperform pure data augmented training. However, \cite{kliegl2017trace} showed that weight decay is most effective with separate $\lambda_{rec}$ and $\lambda_{nonrec}$ hyper-parameters for determining the strength of regularization for the recurrent and non-recurrent weight matrices. We have not tried this in our experiments. Additionally, we discovered that applying multi-condition training while naively lowering the activations of hidden representations leads to nearly identical performance (on both the original and out-of-domain noise perturbed test data) and convergence trajectory as the base model trained on noise augmented data. These results support our hypothesis that models trained with the IRL algorithm do not trivially decrease the magnitude of intermediate representations.

Our final experiments test the effects of various out-of-domain noise on our models. The results are shown in Table \ref{tab:outofdomain}. We found that our models trained with the IRL procedure had stronger results (and significantly less degradation) across all tasks compared to the baseline and the purely data augmented models. When applying various room reverberation on speech, we found that the IRL-C model had a character error rate of $13.8\%$ compared to $21.0\%$ on the data augmented model and $24.1\%$ on the baseline model. Our IRL-C model shows $14.1\%$ character error rate on out-of-domain overlapping speech compared to $91.5\%$ for the baseline and $32.0\%$ on the data augmented model. We found that decreasing the signal-to-noise ratio also effected the baseline models more than the models trained on the IRL algorithm: our IRL-C model received a character error rate of $5.7\%$ compared to $10.8\%$ for baseline and $27.8\%$ for the purely data augmented model. We found that modifying the volume of the speaker did not effect the accuracy of any of the networks. Finally, we found that our models trained with the IRL algorithm performed better for re-sampled telephoney data, achieving a character error rate of $6.4\%$ for IRL-C compared to $14.2\%$ for baseline and $12.2\%$ for the purely data augmented model.

\begin{figure*}
\begin{subfigure}[t]{0.5\textwidth}
  \includegraphics[keepaspectratio, width=1\textwidth]{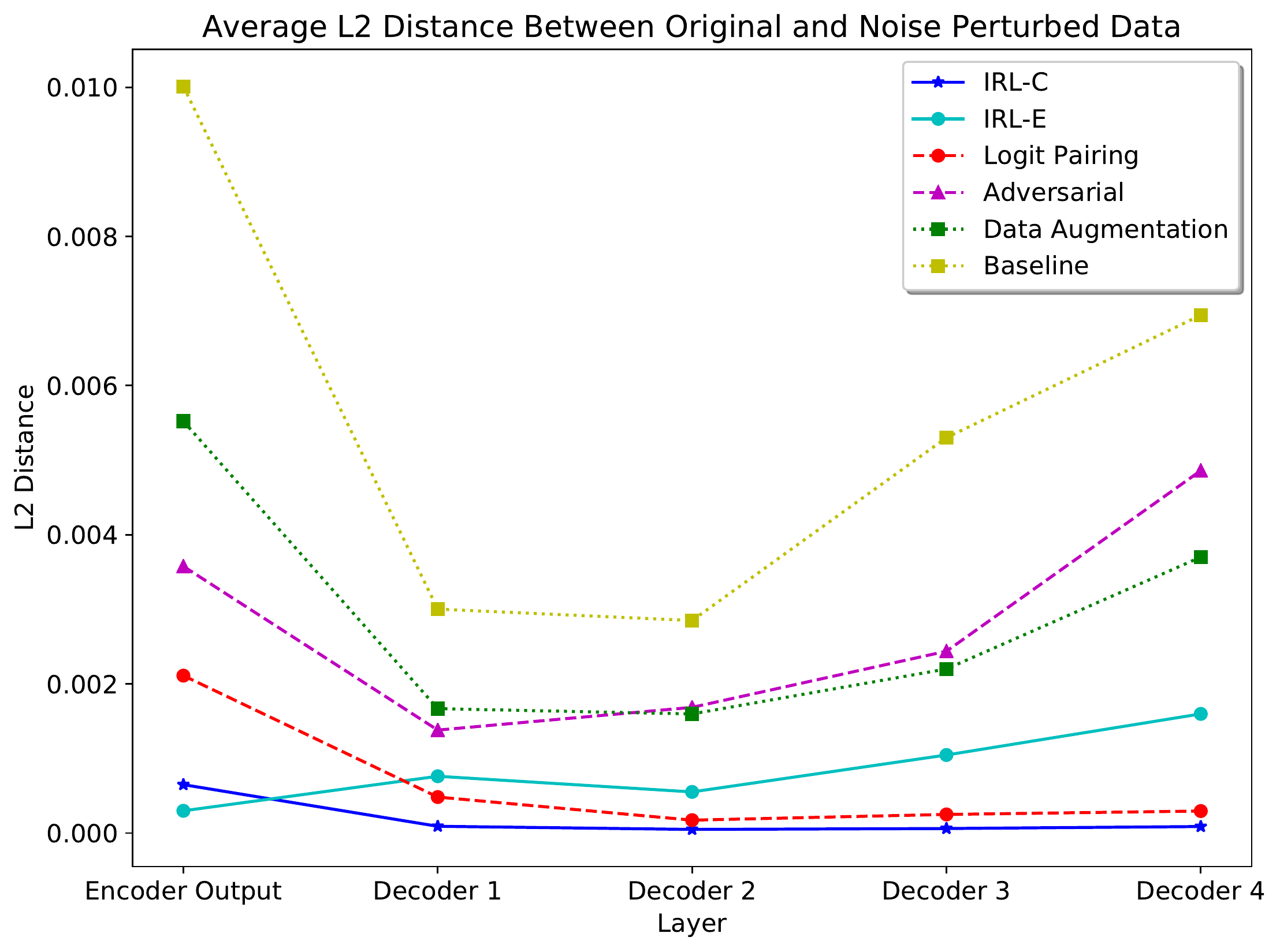}
  \caption{L2 distance}
  \label{fig:l2distance}
\end{subfigure}~
\begin{subfigure}[t]{.5\textwidth}
\centering
  \includegraphics[keepaspectratio, width=1\textwidth]{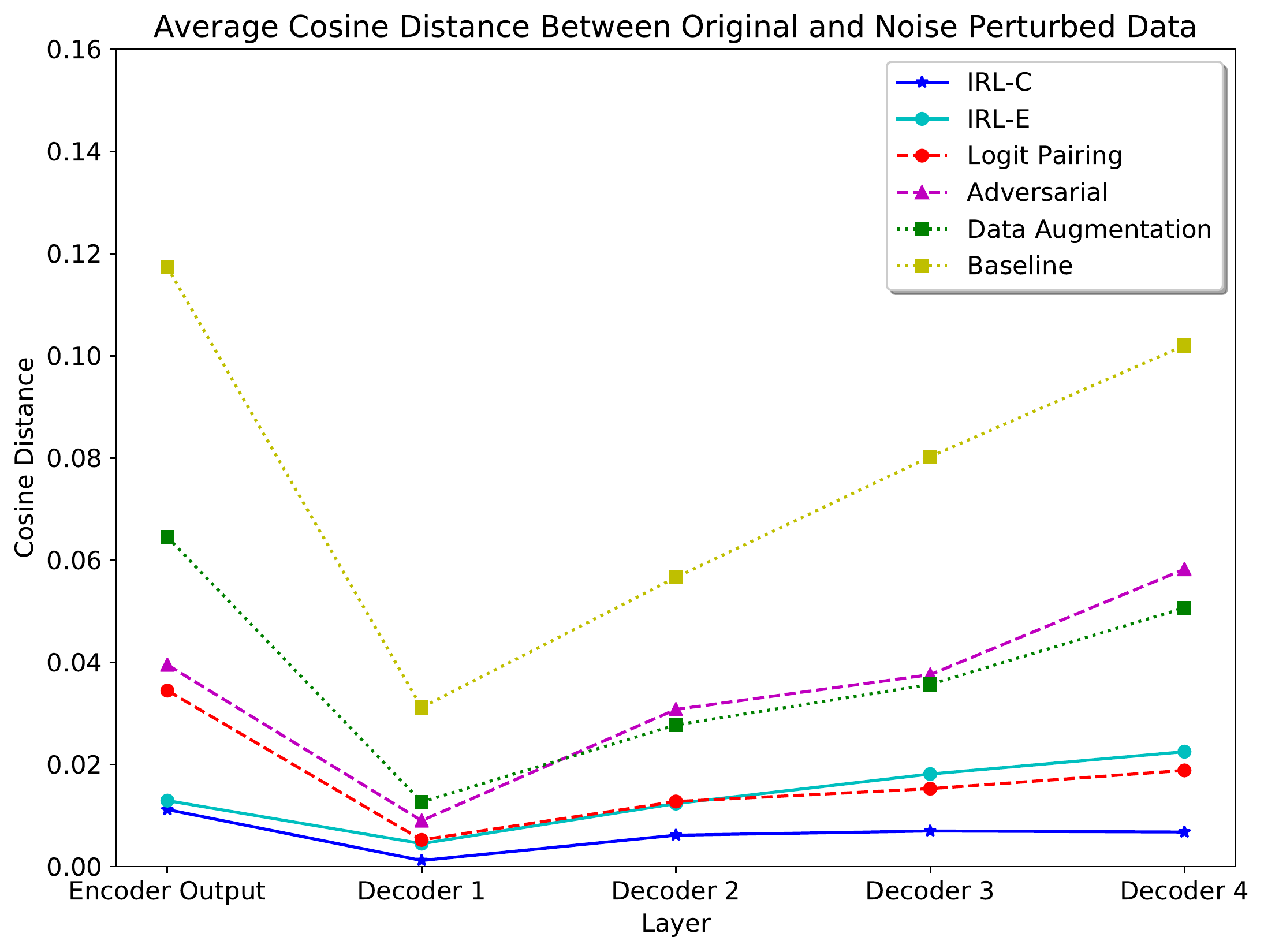}
  \caption{Cosine distance}
  \label{fig:cosinedistance}
\end{subfigure}
\caption{Average distance between original and noised data for various models (distinct lines) and various layers (x-axis). Subplot (a) depicts L2 distance and (b) depicts cosine distance.}
\label{fig:distances}
\end{figure*}

We also executed some empirical analysis
to determine the effect of the various approaches on the distances between noisy examples and their clean counterparts in representation space.
In general, our IRL models have the lowest L2 
and cosine distances between noisy representations and the clean counterparts.
In Figure \ref{fig:distances}, 
you can see that although the IRL-E and IRL-C model models have similarly close representations at the encoder layer, neither reaches $0$ distance.
Then for IRL-E over the subsequent layers,
the clean and noisy representations diverge again,
while for IRL-C they remain close throughout.

\section{Conclusions}
\label{sec:conclusions}
In this paper, we demonstrated 
that enforcing noise-invariant representations 
by penalizing differences between pairs of clean and noisy data 
can increase model accuracy on the ASR task,
produce models that are robust to out-of-domain noise, and improve convergence speed. 
The performance gains achieved by IRL
come without any impact to inference throughput. 
We note that our core ideas here can be applied broadly to deep networks for any supervised task.
While the speech setting is particularly interesting to us, 
our methods are equally applicable to 
other machine learning fields, 
notably computer vision. 
One natural extension might be to experiment with various other loss functions such as triplet losses,
requiring that noisy data be both close to its clean counterpart and further away from \emph{different} clean data.
Additionally, our approach might be well-suited to conferring greater robustness to adversarial examples.  
The comparative improvements over requiring invariant hidden representations vs. invariant logits here raises the possibility 
that we might be able to realize similar gains over logit pairing in the adversarial setting.


\section{Acknowledgements}
\label{sec:acknowledgements}
The authors would like to thank 
Jeremy Cohen,
Mukul Kumar, 
Karishma Malkan, 
Julian Salazar, 
Alex Smola, and
Jerry Zhang
for helpful discussions and suggestions.

\bibliography{main}{}
\bibliographystyle{IEEEbib}

\end{document}